\documentclass[12pt, 
prd, 
nofootinbib,
 amsmath,amssymb,
 aps,
]{revtex4-2}

\usepackage{graphicx}
\usepackage{xcolor}
\usepackage{dcolumn}
\usepackage{bm}
\usepackage{hyperref}
\usepackage{cleveref}
\newcommand{\Yeq}{Y_{\text{eq}}}
\newcommand{\Mp}{M_{\text{pl}}}

\newcommand{\TRH}{T_{\text{RH}}}
\newcommand{\TG}{T_{\rm GUT}}

\begin{document}

\preprint{APS/123-QED}

\title{Reheating Bounds from Thermal GUT Monopole Production}

\author{Donald Liveoak}
\affiliation{
 Leinweber Institute for Theoretical Physics, Physics Department, University of Michigan, Ann Arbor, MI, USA
}

\author{Arielle Schutz}
\affiliation{
 Leinweber Institute for Theoretical Physics, Physics Department, University of Michigan, Ann Arbor, MI, USA
}

\author{James D. Wells}
\affiliation{
 Leinweber Institute for Theoretical Physics, Physics Department, University of Michigan, Ann Arbor, MI, USA
}

\date{\today}

\begin{abstract}
\textit{Abstract:} Magnetic monopoles are a generic prediction of Grand Unified Theories (GUTs) that are in tension with modern cosmological and observational bounds. In this paper, we calculate the present-day abundance of GUT monopoles produced thermally in the early Universe. We improve on previous calculations by accounting for full relativistic corrections to the thermal abundance and enhanced monopole annihilation due to the emission of radiation and scattering off massive gauge bosons, the Standard Model fermions and their superpartners. To obtain a present-day energy density less than that of dark matter, we show that the reheating temperature of the Universe, $\TRH$, must be less than $0.55$ times the GUT symmetry breaking scale $\TG$ for the canonical 't Hooft-Polyakov monopole mass. Accounting for Parker bounds on the present-day magnetic monopole abundance, the bound is tightened to $T_{\text{RH}}/T_{\text{GUT}} \lesssim 0.45$. Furthermore, experimental bounds from Super-Kamiokande require $T_{\text{RH}}/T_{\text{GUT}} \lesssim 0.35$ in our scenario.
\end{abstract}

\maketitle

\section{Introduction}
Grand Unified Theories (GUTs) are a class of well-motivated extensions to the Standard Model (SM) in which the SM gauge group is embedded as a subgroup of a higher-rank Lie group, such as $SU(5)$ or $SO(10)$. GUTs naturally support a variety of beyond the Standard Model (BSM) phenomena, including gauge coupling unification \cite{ellis1991probing}, charge quantization \cite{pati1974lepton}, and the cancellation of gauge anomalies \cite{georgi1974unity}. Furthermore, several GUT models naturally incorporate mechanisms for baryogenesis \cite{yoshimura1978unified, kolb1996grand, hooper2021gut} and the generation of neutrino masses via the seesaw mechanism \cite{minkowski1977mu, gell2010complex}. GUTs are also explicit in heterotic string theory constructions \cite{gross1985heterotic}.

However, despite their theoretical success, experimental and observational signatures of GUT physics remain elusive. Heavy gauge bosons of the broken GUT symmetry mediate the proton decay process $p\to \pi^0 +e^+$, which has yet to be observed. Present-day bounds on the proton lifetime are $\tau_p \gtrsim 2.4\times10^{34} \, \text{years}$ \cite{takenaka2020search}, which rules out a number of minimal GUTs \cite{OHLSSON2023116268, georgi1974unity, dimopoulos1981softly, SAKAI1982533, PhysRevD.32.2348, BABU1998337}.

Additionally, GUTs predict a spectrum of topologically stable magnetic monopole states \cite{t1974magnetic, polyakov1974particle} which may be produced copiously in a phase transition in the early Universe \cite{kibble1976topology, zurek1985cosmological}. During this phase transition, the GUT gauge group $G_{\rm GUT}$ is broken to a subgroup containing the SM gauge group $G_{\rm SM}=SU(3)_C \times SU(2)_L \times U(1)_Y$. After the phase transition, monopoles rapidly overclose the Universe and initiate a period of matter domination, which spoils Big Bang nucleosynthesis (BBN) \cite{zeldovich1978concentration, preskill1979cosmological}. This so-called \textit{monopole problem} was an early motivation for inflation \cite{guth1981inflationary}. If inflation occurred in the early Universe and reheated the Universe to a temperature $\TRH$ much below the symmetry breaking scale $\TG$, GUT monopoles produced at the phase transition may be sufficiently diluted to protect the standard radiation-dominated cosmology.

If the monopole density is effectively diluted by inflation but the reheating temperature is near $\TG$, monopole-antimonopole pairs may be generated thermally from the bath of relativistic charged particles \cite{turner1982thermal}. In this case, monopoles are said to be \textit{frozen in}. This scenario may still lead to an overabundance of monopoles. On the other hand, this mechanism may produce an abundance of magnetic monopoles which is accessible in future observational and experimental efforts. 

Even if the monopole abundance is sufficiently small as to preserve radiation domination at BBN, their historical and present-day abundances are strongly constrained. First, the present-day energy density of monopoles must not exceed that of dark matter (DM). Furthermore, Parker bounds require the flux of magnetic monopoles to be sufficiently small as to not drain Galactic magnetic fields \cite{Parker1970, TurnerParkerBogdan1982}. Finally, the flux of monopoles must be small enough to evade bounds from modern direct-detection experiments, such as MACRO \cite{MACRO_2002}.

In this paper, we carefully calculate the abundance of GUT monopoles produced via freeze-in in the early Universe. We account for thermal production at temperatures near the GUT scale via the collision of relativistic charged particles. We improve on the classical calculation by Turner \cite{turner1982thermal} by accounting for the full relativistic corrections to the equilibrium abundance, as well as BSM degrees of freedom (massive gauge bosons and charged superpartners) which may be abundant in scenarios in which the Universe is reheated near the GUT scale. We find that $T_{\text{RH}}/T_{\text{GUT}} \lesssim 0.55$ is generally required to have a present-day monopole energy density less than that of DM. Furthermore, we find $T_{\text{RH}}/T_{\text{GUT}} \lesssim 0.35$ to evade current observational and experimental bounds.

This paper is structured as follows. In \Cref{sec:mechanisms}, we present the thermal and non-thermal  mechanisms which produce magnetic monopoles in the early Universe. In \Cref{sec:abundance}, we derive an upper bound on the reheating temperature based on the observational constraints, and confirm it numerically. In \Cref{sec:obs}, we apply our calculation to modern observational constraints on the present-day monopole abundance. In \Cref{sec:discussion}, we discuss potential mechanisms for evading our bounds, including cosmic strings and exotic SSB chains. Additionally, we comment on the application of our bounds to intermediate-mass monopoles. We conclude in \Cref{sec:conclusion}.

\section{Monopole evolution}\label{sec:mechanisms}
If a gauge symmetry $G$ is spontaneously broken into a subgroup $H$, topologically stable monopole solutions will be permitted if the second homotopy group of the vacuum manifold, $\pi_2(G/H)$, is nontrivial. This is generally the case in GUTs, since if the SM gauge group $G_{\text{SM}}=SU(3)_C\times SU(2)_L\times U(1)_Y$ is embedded in a simple Lie group $G_{\text{GUT}}$, then
\begin{equation}
    \pi_2(G_{\text{GUT}}/G_{\text{SM}}) \simeq \pi_1(G_{\rm SM}) \simeq \pi_1(SU(3))\oplus \pi_1(SU(2))\oplus \pi_1(U(1)) \simeq \mathbb{Z},
\end{equation}
since $SU(2)$ and $SU(3)$ are simply connected. If $G_{\rm GUT}$ is broken to $G_{\rm SM}$ via a multi-stage symmetry breaking pattern, $H_1 \to H_2 \to \cdots \to  H_n$, monopoles will be formed at any step for which the vacuum manifold $H_k/H_{k+1}$ has a non-trivial second homotopy group. 

\subsection{Monopole production}
We now focus on the case of a single-stage breaking $G_{\text{GUT}} \to G_{\text{SM}}$ which occurs during a thermal phase transition in the early Universe at temperature $T=T_\text{GUT}$. The mass and magnetic charge of monopoles are $m_M =4\pi \kappa_M\TG/g$ and $h=4\pi /g$, where $g$ is the gauge coupling \footnote{Here, we identify the symmetry-breaking scale $\TG$ with the vacuum expectation value; model-dependent deviations between these two parameters may be absorbed into the definition of $\kappa_M$.}. Here, $\kappa_M$ is an $\mathcal{O}(1)$ factor which may depend on group theory factors \citep{weinberg1980fundamental}, the quartic Higgs self-coupling $\lambda_\phi$ \citep{kirkman1981asymptotic} and the details of reheating \cite{collins1984thermal}. For 't Hooft-Polyakov monopoles, the group theory factor is trivial, and $\kappa_M\to 1$ in the Prasad-Sommerfield limit where $\lambda_\phi\to 0$ \cite{prasad1975exact}. Additionally, for supersymmetric GUTs, $g\simeq 1/\sqrt{2}$ near the unification scale. 

During the phase transition, monopoles may be produced at high abundances. If the phase transition is second-order or weakly first-order, the production of monopoles is described by the Kibble-Zurek mechanism. The initial estimate on the number density $n_M$\footnote{We assume that the number density of monopoles and antimonopoles is equal.} of non-thermal monopoles in this scenario was given by Kibble, who argued on causal grounds that $n_M$ must exceed $\sim 1$ per horizon volume $H^{-3}$, with $H$ the Hubble parameter \cite{kibble1976topology}. Zurek later refined this estimate by carefully considering the quenching behavior of the phase transition, finding that the initial number density $n_M H^{-3}\sim (\Mp/\TG)^2$ \cite{zurek1985cosmological}, where $\Mp=2.44\times10^{18}\, \rm GeV$ is the reduced Planck mass. If the phase transition is strongly first-order, monopoles are produced during the collisions of bubbles of true vacuum nucleating into the false vacuum. For mean bubble radius at collision $R_p$, the initial monopole density is $n_M \sim R_p^{-3}$ \cite{brummer2026no}.

After the phase transition, the number density of monopoles is governed by the Boltzmann equation \cite{kolb1981early}
\begin{equation}
    \frac{dn_M}{dt} + 3Hn_M= -D(n_M^2-n^2_{M,\rm eq}(T)),
\end{equation}
where $n_{M,\rm eq}(T)$ is the equilibrium number density of monopoles at temperature $T$, and $D$ is the thermally-averaged cross section of monopole-antimonopole annihilation. Detailed balance enforces $n_M(T)=n_{M,\rm eq}(T)$ at equilibrium. Thus, if the initial monopole density is sub-thermal, monopoles will be thermally produced via inverse annihilation processes. 

\subsection{Monopole annihilation}\label{sec:annihilation}
After monopoles $M$ and antimonopoles $\bar{M}$ form, they may annihilate via $M + \bar{M} \to \gamma \gamma$ or similar processes.

First, magnetic monopoles may annihilate via diffusive capture in which monopole-antimonopole pairs diffuse via interactions with the plasma of relativistic charged particles in which they are immersed. In this process, monopoles lose energy via scattering with the plasma and are eventually captured into monopole-antimonopole Coulomb bound states which cascade and annihilate \cite{preskill1979cosmological}. This yields the annihilation cross section
\begin{equation}
    D_\text{diffuse} =(h^2/B)m_M^{-2} (m_M/T)^2, \label{eq:D_Coulomb}
\end{equation}
where
\begin{equation}\label{eq:B}
    B \simeq \frac{2\pi}{9} \sum_{i} b_i \left(\frac{he_i}{4\pi}\right)^2 \ln (\Lambda) 
\end{equation}
is the drag coefficient of the monopoles, which enumerates the number of spin states of charged relativistic particles \cite{dunsky2022guts, vilenkin1994cosmic}. Here $b_i=1$ for bosons and $b_i=1/2$ for fermions, $e_i$ is the charge of the $i^{\text{th}}$ spin state and $\Lambda\sim (16\pi^2/g_\ast g^4)$ is the ratio between the maximum and minimum scattering angles of the monopoles, with $g_\ast$ the effective number of relativistic spin degrees of freedom ($1$ per bosonic d.o.f., $7/8$ per fermionic d.o.f.) \cite{vilenkin1994cosmic}.

For the charged fermions of the SM, we have $B\approx 20$. However, at the GUT scale, we expect the plasma to contain additional relativistic degrees of freedom, including superpartners and charged $X$ and $Y$ bosons. Though the $X$ and $Y$ bosons are unstable and may decay into quarks and leptons, we expect them to maintain their equilibrium abundance via inverse decay when $T$ exceeds the mass of the gauge bosons $M_X\sim g\TG$. Thus, in our analysis, we adopt 
\begin{equation}
    B = B_{\text{SM}} + B_{\text{SUSY}} + B_{XY}\theta(T-g\TG),
\end{equation}
where $B_{\rm SM} \approx 20$ counts the Standard Model fermions, $B_{\rm SUSY}$ accounts for the charged superpartners, and $B_{XY}$ accounts for charged gauge bosons. The Heaviside function $\theta(T-gT_\text{GUT})$ arises since the abundance of gauge bosons becomes exponentially suppressed once $T < M_{X} \sim gT_{\rm GUT}$. We adopt $B_{\rm SUSY}=2B_{\rm SM}=40$, since each charged fermion has a charged sfermion partner with the same quantum numbers but $b_i=1$ instead of $b_i=1/2$ in \Cref{eq:B}. Furthermore, the precise value of $B_{XY}$ depends on the spectrum of massive gauge bosons, which in turn depends on the rank of the unbroken gauge group. Here, we choose minimal $SU(5)$, which yields $B_{XY}\approx 31$.

Annihilation due to interaction with the charged plasma is efficient until the mean free path of the monopoles exceeds the Coulomb capture distance at $T_{\rm stop}=(4\pi/h^2)^2 m_M B^{-2}$ \cite{preskill1979cosmological}. Below $T_{\rm stop}$, $M-\bar{M}$ annihilation may occur via the emission of radiation (bremsstrahlung). This leads to a substantially weaker annihilation cross section \cite{Elyutin:1978recombination}
\begin{equation}
    D_{\rm brem} = (h^2/4\pi)^2 m_M^{-2} (m_M/T)^{9/10}.\label{eq:D_brem}
\end{equation}
This annihilation is cut off by the expansion of the Universe when the interaction rate $\Gamma_{\text{brem}}=D_{\rm brem} n_M < H$.

\subsection{Inflation, reheating, and monopole problems}
When monopoles are produced via a phase transition, annihilation is generally insufficient to dilute them to low abundances \cite{preskill1979cosmological}. In this case, the monopoles quickly overclose the Universe, initiating a period of matter domination before BBN. To avoid this monopole problem, Guth proposed a period of cosmic inflation at temperatures below $\TG$, in which the monopole density is exponentially diluted to acceptable values \cite{guth1981inflationary}. After inflation, the energy density of the Universe is transferred into Standard Model particles in a process known as reheating, at which point the Universe equilibrates at a temperature $\TRH$.

If $\TRH > \TG$, the GUT symmetry is restored. As the Universe cools, the symmetry will once again be spontaneously broken, again populating the Universe with monopoles and leading to an overclosure problem. If $\TRH \ll \TG$, the monopoles will remain at effectively zero abundance, preserving the standard cosmology. In the intermediate case that $\TRH \lesssim \TG$ but not by much, monopoles will be thermally produced and either lead to an unacceptably high abundance, negligible abundance, or an intermediate abundance potentially accessible to modern experimental and observational efforts.

\section{Monopole relic abundance}\label{sec:abundance}

In this section, we analyze the evolution of the monopole density via annihilation and reverse annihilation, as described in Section \ref{sec:annihilation}. First, we write analytic expressions for a general annihilation coefficient. Then, we present the results of our numerical integrations of the Boltzmann equation and the resulting bounds on $\TRH/\TG$.

\subsection{Analytical expectations}\label{sec:anal}
We consider the monopole abundance analytically for the case of an annihilation coefficient of the form $D =(A/m_M^2)x^2$, where $x\equiv m_M/T$. In the case of diffusive capture, $A = h^2/B$. In this analysis, we allow $A$ and the initial monopole density $n_M^i$ to vary and define the normalized initial monopole number density
\begin{equation}
    \alpha \equiv \frac{n_M^i}{\TRH^3}\frac{AC\Mp}{m_M},
\end{equation}
where $\TRH$ is the temperature at reheating, and $C=\sqrt{90/\pi^2 g_\ast}$. Then, the Boltzmann equation (neglecting thermal production) integrates to
\begin{equation}
    \frac{n_M}{T^3} = \left(\frac{\TRH^3}{n_M^i} + \frac{AC\Mp}{m_M}\left(x-x_i\right)\right)^{-1}=\frac{m_M}{AC\Mp}\left(\frac{1}{\alpha} + (x-x_i)\right)^{-1}.
\end{equation}
In the limit $\alpha \to 0$, we recover the constant entropy density solution $n_M/T^3=n_M^i/\TRH^3$. 

Additionally, we define the normalized initial temperature $\beta \equiv \TRH/\TG$. In the case where $\beta > 1$, copious monopoles are produced via the Kibble-Zurek mechanism at the phase transition $T=\TG$; for $\beta < 1$, the dominant production mechanism is thermal production via the inverse annihilation mechanism $\gamma\gamma \to M+\bar{M}$ (and similar mechanisms for charged particles).

It is helpful to cast the Boltzmann equation in terms of the comoving abundance of monopoles, the yield $Y_M\equiv n_M/s$, where $s= (2\pi^2/45)g_\ast T^3$ is the entropy density. In this analysis, we assume that all light particle species remain relativistic during the integration period, so that $g_\ast$ remains constant.

The Boltzmann equation (including the thermal production term) is
\begin{equation}\label{eq:boltzmann}
    \frac{dY_M}{dx} = -\frac{\lambda D}{x^2} (Y_M^2 - \Yeq^2),
\end{equation}
where we define
\begin{equation}
    \lambda = \frac{sx}{H} = \sqrt{\frac{90}{\pi^2 g_\ast}} \frac{2\pi^2}{45} g_\ast \Mp m_M,
\end{equation}
and the equilibrium abundance $\Yeq(x)\simeq (45/4\pi^4g_\ast) x^2K_2(x)$, with $K_2$ the modified Bessel function of the second kind. In the non-relativistic limit $x\gg1$, $\Yeq(x)\simeq (0.145/g_\ast)x^{3/2} e^{-x}$.

In the case of thermal production, we assume the initial monopole density is negligible so that $\alpha \ll 1$. Thus, we have that $Y_M \ll \Yeq$ so annihilation cannot meaningfully decrease the monopole abundance. We may then estimate the final abundance of monopoles by integrating the Boltzmann equation
\begin{equation}
    \frac{dY_M}{dx} \approx \frac{\lambda D}{x^2}Y_\text{eq}(x)^2.
\end{equation}
 If the initial temperature is sufficiently close to the critical temperature (i.e. $\beta \sim 1$), $\frac{\lambda D}{x^2}\Yeq(x)^2$ may be non-trivial and lead to an abundance of monopoles. Thus, we estimate the asymptotic abundance, $Y_\infty$, as
\begin{equation}
    Y_\infty = Y_i+\int_{x_i}^{x_f}dx\,\frac{dY_M}{dx} \approx \alpha \left(\frac{AC\Mp}{m_M}\right)^{-1} \frac{45}{2\pi^2} \frac{1}{g_\ast} + \int_{x_i=4\pi \kappa_M/g\beta }^\infty \frac{\lambda D}{x^2} \Yeq(x)^2\,dx,
\end{equation}
where $Y_i$ is the initial comoving abundance of monopoles. Since $4\pi \kappa_M /g\beta \gg 1$, we take the non-relativistic ($x\gg1$) limit of $\Yeq(x)$ to find
\begin{equation}
    \int_{4\pi \kappa_M/g\beta}^\infty \frac{\lambda D}{x^2} \Yeq(x)^2 \approx \left(\frac{0.145}{g_\ast}\right)^2 \frac{\lambda A}{m_M^2} \int_{4\pi \kappa_M / g\beta }^\infty x^3\, e^{-2x}\, dx \equiv \Delta Y_M(\beta),
\end{equation}
which is computable exactly via integration by parts. Thus, we may compute
\begin{equation}
    Y_\infty \approx Y_i(\alpha) + \Delta Y_M(\beta),
\end{equation}
analytically. At the present epoch, the fractional energy of monopoles $\Omega_M$ must not exceed the energy density of dark matter $\Omega_{\text{DM}}$:
\begin{equation}
    \Omega_M \equiv \frac{Y_\infty s_0m_M}{\rho_c} < \Omega_{\text{DM}} \sim 0.25.
\end{equation}
Here, $s_0$ and $\rho_c$ are the present-day values of the entropy density and critical energy density
\begin{align}
    s_0 = \frac{43}{11}\frac{2\pi^2}{45}T_0^3 &\sim 2.2\times 10^{-38}\, \text{GeV}^3,\\
    \rho_c = 3\Mp^2H_0^2 &\sim 3.6\times 10^{-47}\, \text{GeV}^4,
\label{eq:numerical_params}
\end{align}
where $H_0$ is the Hubble constant and $T_0$ is the CMB temperature. This immediately implies
\begin{equation}
    \alpha < \frac{2\pi^2}{45}\frac{AC\Mp g_\ast}{m_M} \frac{\rho_c\Omega_{\rm DM}}{s_0m_M}.
\end{equation}
We now derive the upper bound on $\beta$. We consider the case where inflation dilutes the monopoles to a negligible initial density, and thus $\alpha = 0$. Then, to avoid overclosure, we require
\begin{equation}
    Y_\infty = \frac{\lambda A}{m_M^2} \int_{4\pi \kappa_M/ g\beta }^\infty x^3 e^{-2x}\, dx < \frac{\rho_c \Omega_{\rm DM}}{s_0 m_M},
\end{equation}
for $\Omega < \Omega_{\rm DM}$. Noting $\beta < 1$ and thus $x_i=4\pi \kappa_M /g\beta \gg 1$, the integral is
\begin{align}
    Y_\infty &= \frac{1}{8} \left(\frac{0.145}{g_\ast}\right)^2 \frac{A}{m_M^2} \sqrt{\frac{90}{\pi^2 g_\ast}} \frac{2\pi^2}{45} g_\ast \Mp m_M \left(4x_i^3+6x_i^2+6x_i+3\right)\exp({-2x_i})\\
    &\approx \frac{1}{2} \left(\frac{0.145}{g_\ast}\right)^2 A \sqrt{\frac{90}{\pi^2 g_\ast}} \frac{2\pi^2}{45} g_\ast \frac{\Mp}{m_M} \left(\frac{4\pi\kappa_M}{g\beta}\right)^3 \exp({-8\pi \kappa_M / g\beta})\\
    &< \frac{\rho_c \Omega_{\rm DM}}{s_0m_M}.
\end{align}
Hence, we obtain the bound
\begin{align}\label{eq:bound}
    \frac{\beta^3}{\kappa_M^3} \exp(8\pi \kappa_M /g\beta) &> \frac{1}{2} \left(\frac{0.145}{g_\ast}\right)^2 A \sqrt{\frac{90}{\pi^2 g_\ast}} \frac{2\pi^2}{45} g_\ast \Mp \left(\frac{4\pi}{g}\right)^3 \frac{s_0}{\rho_c \Omega_{\text{DM}}}\\
    &= 2\times 10^{25}\times  g_\ast^{-3/2} \left(\frac{4\pi}{g}\right)^3 \frac{A}{\Omega_{\rm DM}}.
\end{align}
The stiffness of $(\beta/\kappa_M)^3 e^{8\kappa_M\pi/g\beta}$ for typical parameter values implies that the bound depends weakly on the spectra of particles in the thermal bath. However, \Cref{eq:bound} is very sensitive to changes in $\kappa_M$; this is intuitive, since a slightly larger value of $\kappa_M$ manifests exponentially in the Boltzmann suppression of $n_M$. We will verify this in \Cref{sec:num}.

For $\Omega_{\text{DM}}h^2=0.12$ \cite{planck2020planck} and fiducial values $\kappa_M=1$, $g=1/\sqrt{2}$, $g_\ast=100-400$ and taking $A=h^2/B_{\rm SM}$, our bound yields $\beta<0.55-0.57$. Accounting for charged superpartners by taking $A=h^2/(B_{\rm SM} + B_{\rm SUSY})$, we have $\beta \lesssim 0.56-0.58$. Thus, we confirm that $\beta \lesssim 0.55$ depends very weakly on the spectrum of charged particles, and requires $\TRH \lesssim 0.55\, \TG$ to avoid overclosure. In the subsequent analysis, we adopt a fiducial value of $g_\ast = 200$ unless otherwise specified.

\begin{figure}
    \centering
    \includegraphics[width=\linewidth]{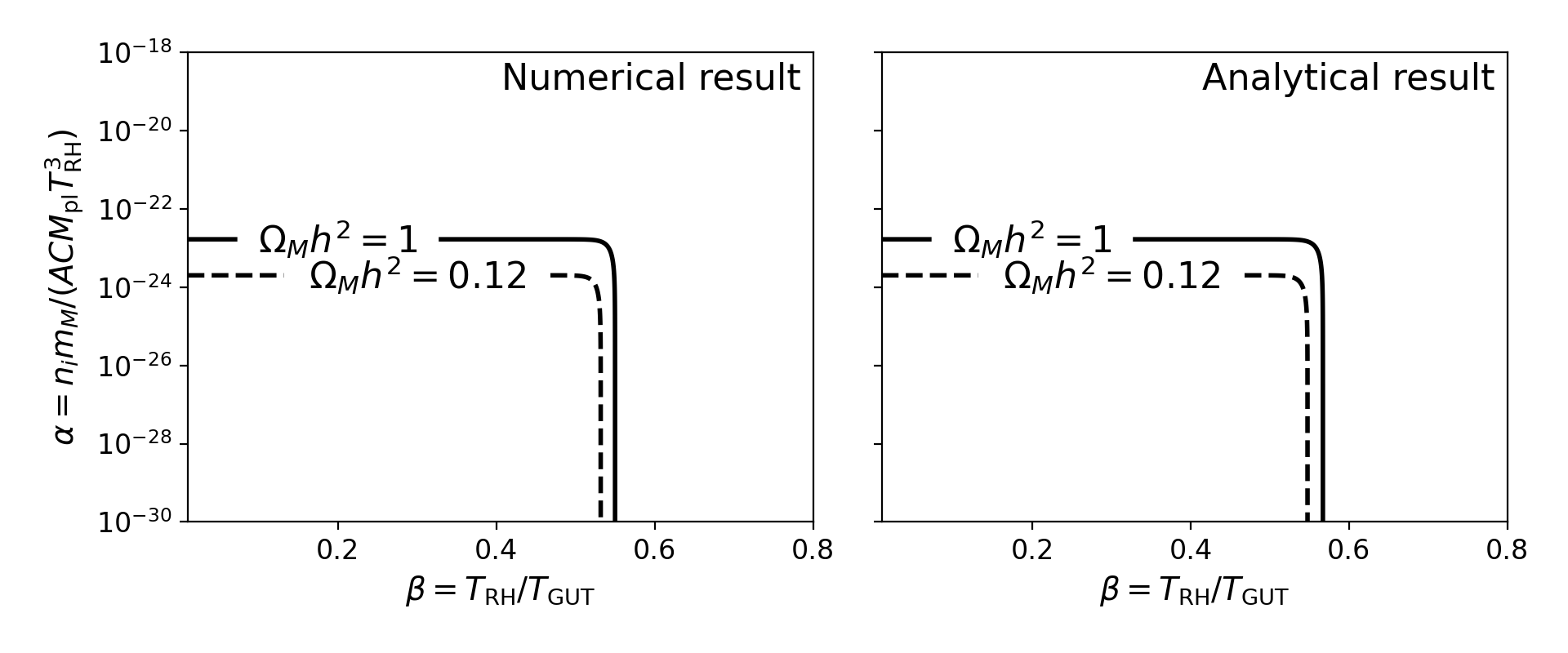}
    \caption{Numerical (left panel) and analytical (right panel) overclosure bounds for thermal magnetic monopoles for $\TG = 10^{16}\, \text{GeV}$ as a function of the normalized initial monopole number density $\alpha$ and normalized reheating temperature $\beta$. The solid and dashed lines indicate $\Omega_M=1$ and $\Omega_M=0.1$, respectively.}
    \label{fig:numerical-analytical-bounds}
\end{figure}

\subsection{Numerical integration}\label{sec:num}
In order to take into consideration the relativistic correction to the equilibrium abundance and the effect of bremsstrahlung-type $M-\bar{M}$ annihilation as described by \Cref{eq:D_brem}, we carry out numerical integrations of the Boltzmann equation. To maintain numerical precision despite $Y$ ranging several orders of magnitude, we use an implicit Runge-Kutta integrator of order $5$. We choose the annihilation cross section
\begin{equation}
    D = D_\text{diffuse} + D_\text{brem} = \frac{h^2 m_M^{-2} x^2}{B
    }\theta(T-m_M B^{-2}(4\pi/h^2)^2) + \left(\frac{h^2}{4\pi}\right)^2 m_M^{-2} 
    x^{9/10} ,
\end{equation}
incorporating both monopole-antimonopole capture due to the scattering of monopoles from the thermal plasma of relativistic charged particles as in \Cref{eq:D_Coulomb}, and the emission of bremsstrahlung as in \Cref{eq:D_brem}. The Heaviside function $\theta(T-m_MB^{-2}(4\pi/h^2)^2)$ enforces that diffusive annihilation is cut off when the mean free path of the monopoles exceeds the Coulomb capture radius.

Motivated by our analytical results, we choose $\log_{10}(\alpha) \in [-30, -20]$ and $\beta \in [0.01, 0.7]$. We integrate from $x_i = m_M/\TRH= 4\pi /g\beta$ until $Y_M(x)$ becomes constant; this typically occurs before $x \sim 100$.

The boundaries $\Omega_Mh^2=0.12$ and $\Omega_Mh^2 =1$ are shown in \Cref{fig:numerical-analytical-bounds} for both our numerical and analytical prescriptions, assuming $\TG=10^{16}\, \text{GeV}$. The bounds generally agree that $\beta \lesssim 0.55$ is required to avoid overclosure. 

\begin{figure}
    \centering
    \includegraphics[width=\linewidth]{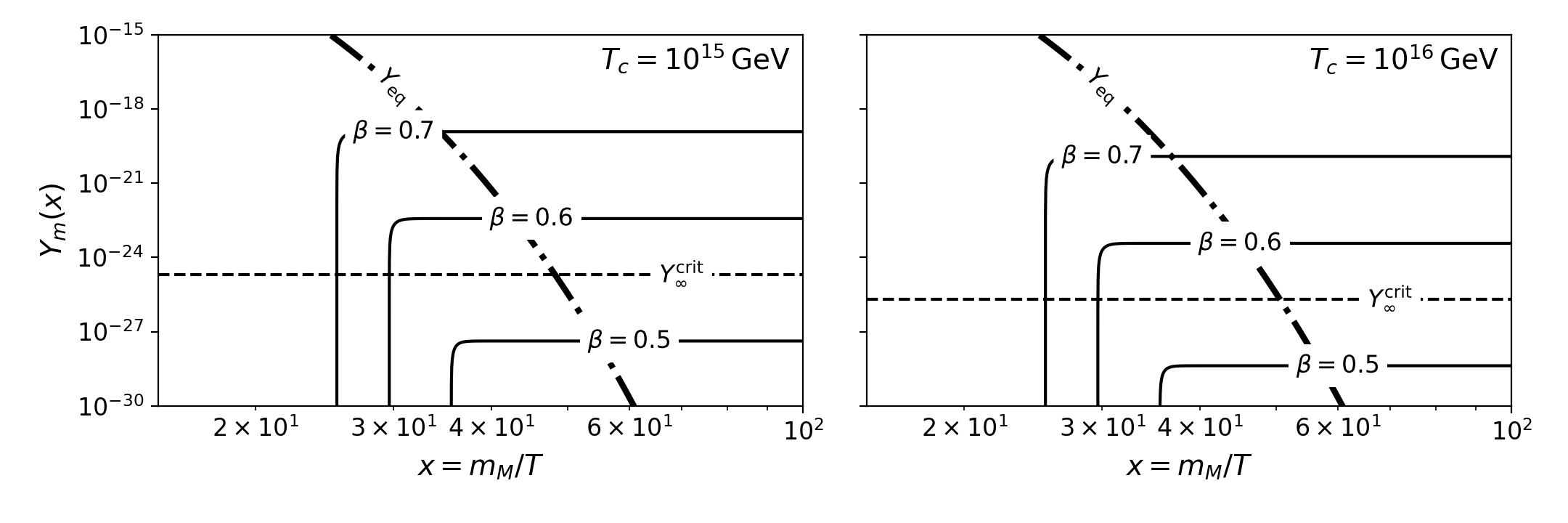}
    \caption{Evolution of the comoving abundance of monopoles, $Y_M(x)$, for $\TG=10^{15}\, \text{GeV}$ (left panel) and $\TG=10^{16} \, \text{GeV}$ (right panel). The dashed line indicates the critical comoving abundance for overclosure, $Y^{\text{crit}}_\infty=\rho_c/s_0 m_M$.}
    \label{fig:timeseries}
\end{figure}

Additionally, for $\TG =10^{15} \, \text{GeV}$ and $\TG = 10^{16}\, \text{GeV}$, we numerically integrate the Boltzmann equation for $\beta=0.4,0.5$, and $0.6$. Again, we choose $\alpha=0$ so that initially, no monopoles are present. The evolution of the comoving abundance $Y_M(x)$ is shown in \Cref{fig:timeseries} for each integration. In each case, we see that $Y_M$ quickly approaches its asymptotic value. For both critical temperatures, $\beta=0.5$ avoids monopole overclosure.

\begin{figure}
    \centering
    \includegraphics[width=0.8\linewidth]{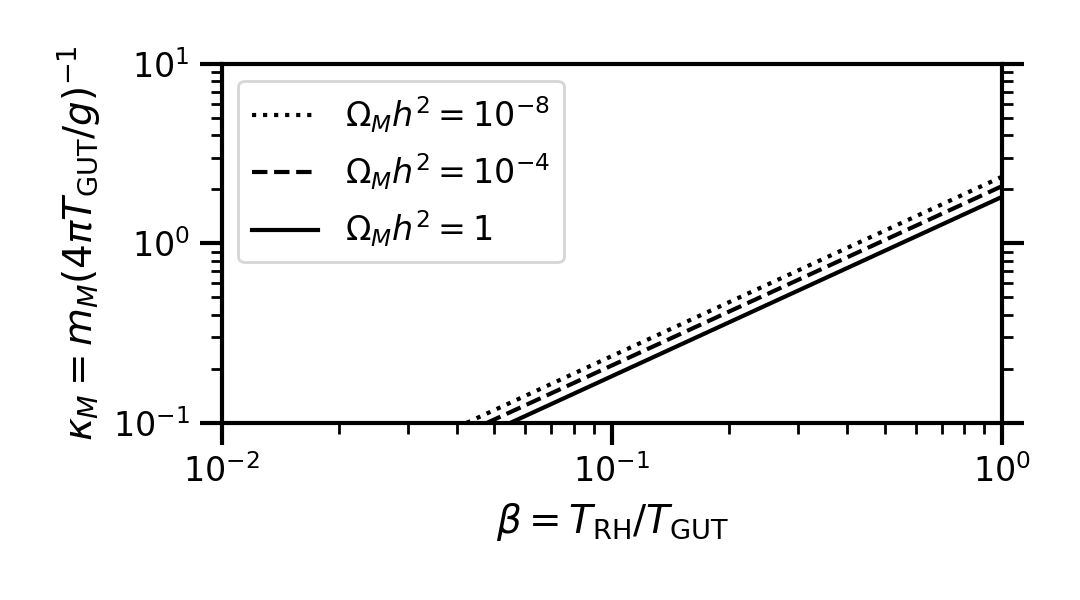}
    \caption{Monopole energy densities $\Omega_M=10^{-8}, 10^{-4},$ and $\Omega_M=1$ as a function of the monopole mass parameter $\kappa_M=m_M(4\pi \TG/g)^{-1}$ and $\beta=\TRH/\TG$. For $\kappa_M \gtrsim 2$, overclosure is avoided, even for reheating near the symmetry restoration scale, $\beta \sim 1$.}
    \label{fig:mass-param}
\end{figure}

Finally, we carry out a set of numerical integrations to determine the dependence of our overclosure boundary on the monopole mass parameter, $\kappa_M=m_M(4\pi \TG/g)^{-1}$. The analytical discussion of the previous section suggests that our bound depends strongly on the choice of $\kappa_M$, due to exponential Boltzmann suppression. Non-standard values of $\kappa_M$ may greatly enhance or suppress the thermal production of monopoles. As such, we integrate the Boltzmann equations, fixing $g=1/\sqrt{2}$, $g_\ast=106.75$, and varying $\kappa\in [0.1, 10]$ and $\beta\in[0.01, 1)$; we note that $\beta=1$ corresponds to symmetry restoration, which would generate an unacceptably high monopole abundance via the Kibble-Zurek mechanism. We plot the contours $\Omega_M(\beta,\kappa_M)=10^{-8},10^{-4}$, and $1$ in \Cref{fig:mass-param} and confirm a sharp dependence of the reheating bound on $\kappa_M$. Specifically, for $\kappa_M \lesssim 0.2$, even reheating to $\TRH \sim 0.1\, \TG$ leads to an overabundance of thermal monopoles. On the other hand, for $\kappa_M \gtrsim 2$, the energy density of thermal monopoles does not exceed that of DM, even for reheating temperatures close to the symmetry-breaking scale. 

Our numerical results confirm and refine the predictions of our analytical calculations in the previous section, which were independent of critical temperature $\TG$ (and thus monopole mass $m_M = 4\pi \kappa_M\TG/g$).  We see that the effect of bremsstrahlung capture is dominated by that of the Coulomb capture, as our numerical results confirm that $\beta \gtrsim 0.55$ generally leads to overclosure. We confirm that our bounds do not depend strongly on $\TG$, nor the spectrum of particles in the thermal bath (quantified by $B$ and $g_\ast$). As expected, our results depend strongly on the ratio of the monopole mass to the symmetry-breaking scale.

\section{Observational constraints on magnetic monopoles}\label{sec:obs}

In this section, we apply our derived overclosure bounds on the energy density of monopoles to modern observational and experimental constraints. A good review of these constraints is available in \cite{Zhang_2024}, which we will closely follow. The first bound is based on the requirement that the mass density of monopoles cannot exceed that of dark matter, as discussed in Section \ref{sec:anal}. This leads to $\Omega_M h^2 < 0.12$ \cite{planck2020planck}. In this section, we discuss two other bounds: the Parker bound, which is related to the interaction of monopoles with the Galactic magnetic field, and bounds from previous and future direct searches, such as MACRO, Super-Kamiokande, and Hyper-Kamiokande. We adopt the canonical value for the monopole mass parameter $\kappa_M=1$.

\subsection{Flux, (anti)monopole velocity, and monopole mass}\label{sec:F_v_m}

First, we note that Parker bounds and bounds from direct searches are often expressed in terms of the limits on magnetic flux $F_M\equiv \frac{n_M v_M}{4\pi}$, where $v_M$ is the typical monopole velocity. Thus, for direct comparison with our analytical and numerical results, we must translate bounds on monopole flux into bounds on their energy density, $\Omega_M$. 

To do this, we must make assumptions regarding the clustering dynamics of magnetic monopoles. This is quantified via the monopole overdensity parameter
\begin{equation}
\Delta = \rho_M/\bar{\rho}_M
\end{equation}
where $\rho_M$ is the local energy density of monopoles where the flux is evaluated, and $\bar{\rho}_M$ is the average energy density of monopoles in the Universe. In the limit of uniformly distributed monopoles, $\Delta=1$. The fractional energy density of monopoles is
\begin{equation}
    \Omega_M = \frac{4\pi m_M}{v_M\rho_c\Delta}F_M.
\end{equation}

Both the monopole velocity $v_M$ and overdensity $\Delta$ depend on monopole mass $m_M$. Heavy monopoles ($\gtrsim 10^{14}\, \rm GeV$) are generally accelerated to modest velocities $v_M\simeq 10^{-3}c-10^{-2}c$ via interactions with Galactic magnetic fields \cite{Zhang_2024,Particle_Data_Review_2022,Medvedev_2017}. The specific value of $\Delta$ may range from 1 to $\sim 10^5$, depending on whether or not monopoles are accelerated to high enough speeds to become gravitationally unbound from galaxies \cite{Zhang_2024}. We adopt the fiducial values $\Delta = 1$ (uniform density) and $\Delta=2.86\times 10^5$ (based on the observed DM overdensity in Milky Way \cite{sofue2020rotation}).

\subsection{Parker bound}\label{sec:parker}
One constraint on the present-day monopole flux is the so-called \textit{Parker bound} \cite{Parker1970}, which requires that the abundance is sufficiently low as not to drain Galactic magnetic fields of their energy. The bound was refined and shown to be mass-dependent by Turner et al. \cite{turner1982magnetic}, and can be written as \cite{Zhang_2024}
\begin{equation}
    F_M < \begin{cases}
    10^{-15}\, \rm cm^{-2} s^{-1} sr^{-1}, & m_M \leq 10^{17}\, \text{GeV} (v_M/10^{-3})^2\\
    10^{-14}\, \text{cm}^{-2} \text{s}^{-1} \text{sr}^{-1} (m_M/10^{18}\, \text{GeV})(v_M/10^{-3})^2, & m_M > 10^{17}\, \text{GeV} (v_M/10^{-3})^2\\
    \end{cases}
\end{equation}
for fiducial values of the Milky Way's magnetic field. This bound is extended by considering the magnetic field of Andromeda; for fiducial parameters, this yields \cite{Zhang_2024}
\begin{equation}
    F_M < \begin{cases}
    5.3\times 10^{-19}\, \rm cm^{-2} s^{-1} sr^{-1}, & m_M \leq 5.3 \times 10^{17}\, \text{GeV} (v_M/10^{-3})^2\\
    10^{-18}\, \text{cm}^{-2} \text{s}^{-1} \text{sr}^{-1} (m_M/10^{18}\, \text{GeV})(v_M/10^{-3})^2, & m_M > 5.3\times 10^{17}\, \text{GeV} (v_M/10^{-3})^2.\\
    \end{cases}
\end{equation}

\begin{figure}
    \centering
    \includegraphics[width=\linewidth]{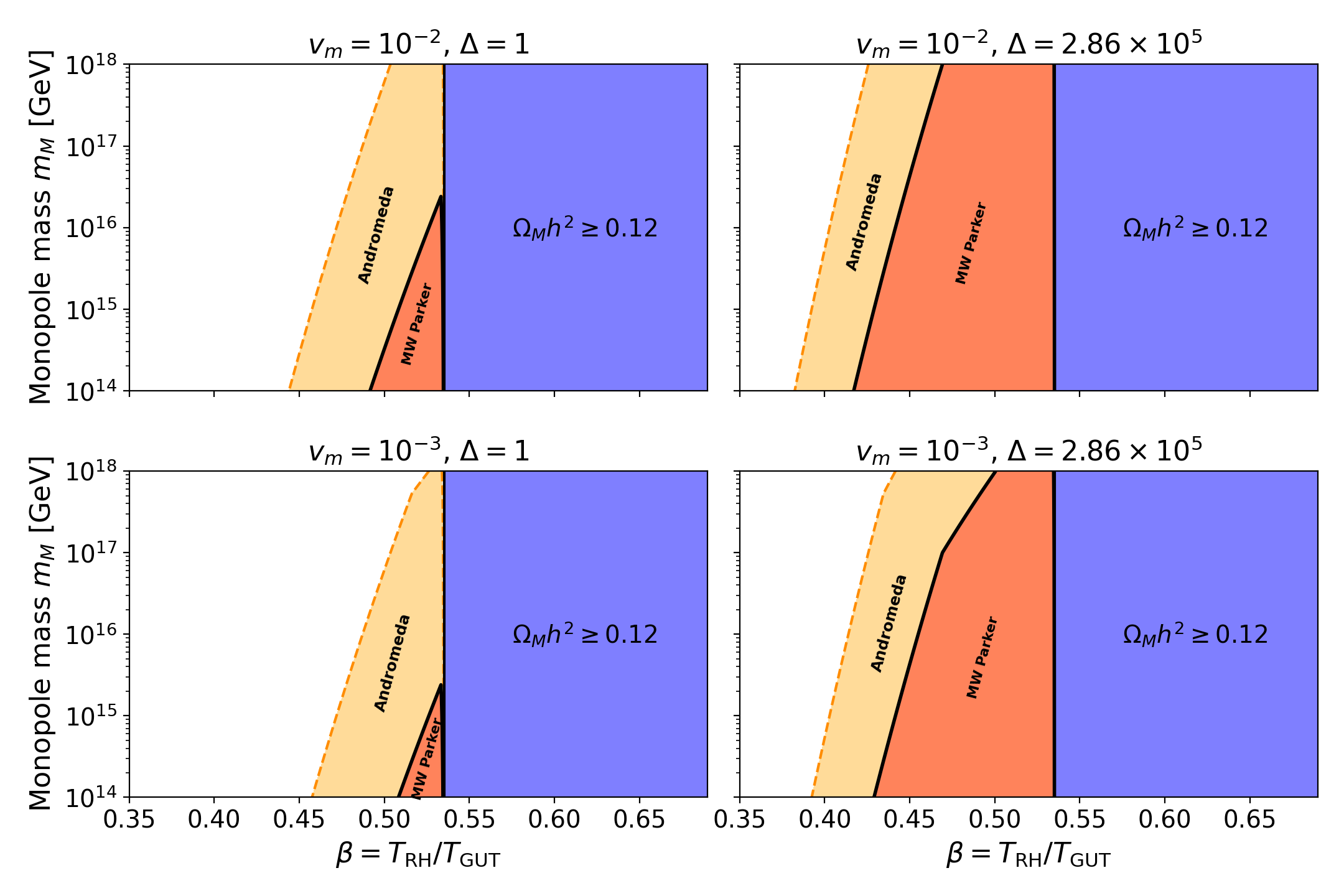}
    \caption{Parker and DM bounds on monopole abundance as a function of the monopole mass $m_M$ and the normalized reheating temperature $\beta$, for various values of the monopole velocity $v_M$ and overdensity $\Delta$.}
    \label{fig:beta-obs-parker}
\end{figure}

\begin{figure}
    \centering
    \includegraphics[width=\linewidth]{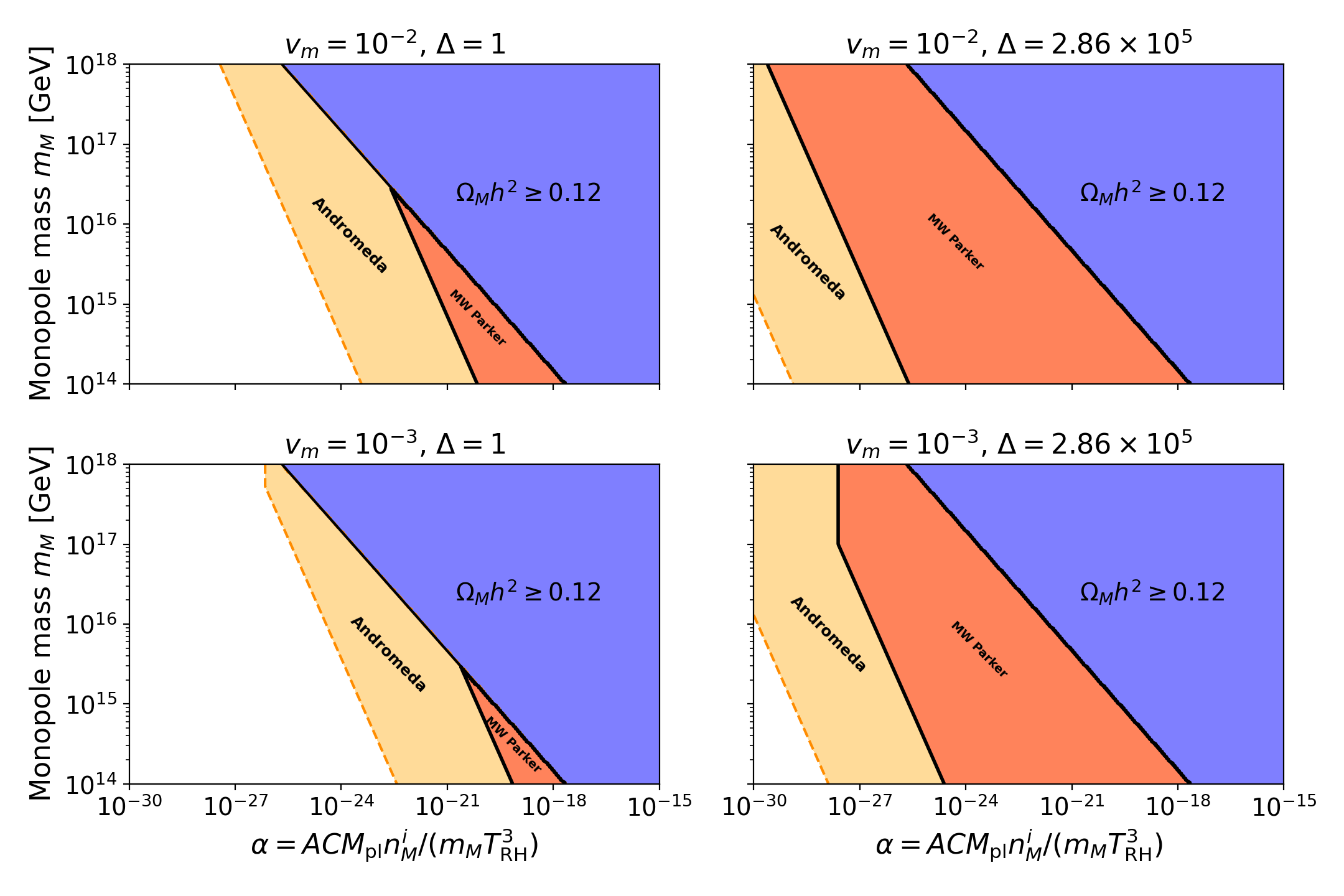}
    \caption{Parker and DM bounds on monopole abundance as a function of the monopole mass $m_M$ and the normalized initial monopole density $\alpha$, for various values of the monopole velocity $v_M$ and overdensity $\Delta$.}
    \label{fig:alpha-obs-parker}
\end{figure}
We now compare these bounds to the results of the numerical integration of the Boltzmann equation. The resulting bounds on the normalized reheating temperature $\beta$ and normalized initial monopole density $\alpha$ are shown in \Cref{fig:beta-obs-parker} and \Cref{fig:alpha-obs-parker} respectively for the fiducial values of the monopole velocity ($v_M=10^{-2},10^{-3}$) and overdensity ($\Delta=1,2.86\times 10^5$). The most extreme case is that of fast ($v_M=10^{-2}$) and clustered ($\Delta=2.86\times 10^5$) monopoles, in which $\beta \lesssim 0.4$ to preserve the Galactic magnetic field at $m_M\sim 10^{14}\, \text{GeV}$. The reheating bound is relaxed to $\beta \lesssim 0.45$ for superheavy monopoles with $m_M \sim 10^{18}\, \text{GeV}$. The case of slower monopoles ($v_M=10^{-3}$) is qualitatively similar but slightly relaxed in the case of superheavy monopoles.

\subsection{Direct detection}\label{sec:direct-detection}
\begin{figure}
    \centering
    \includegraphics[width=\linewidth]{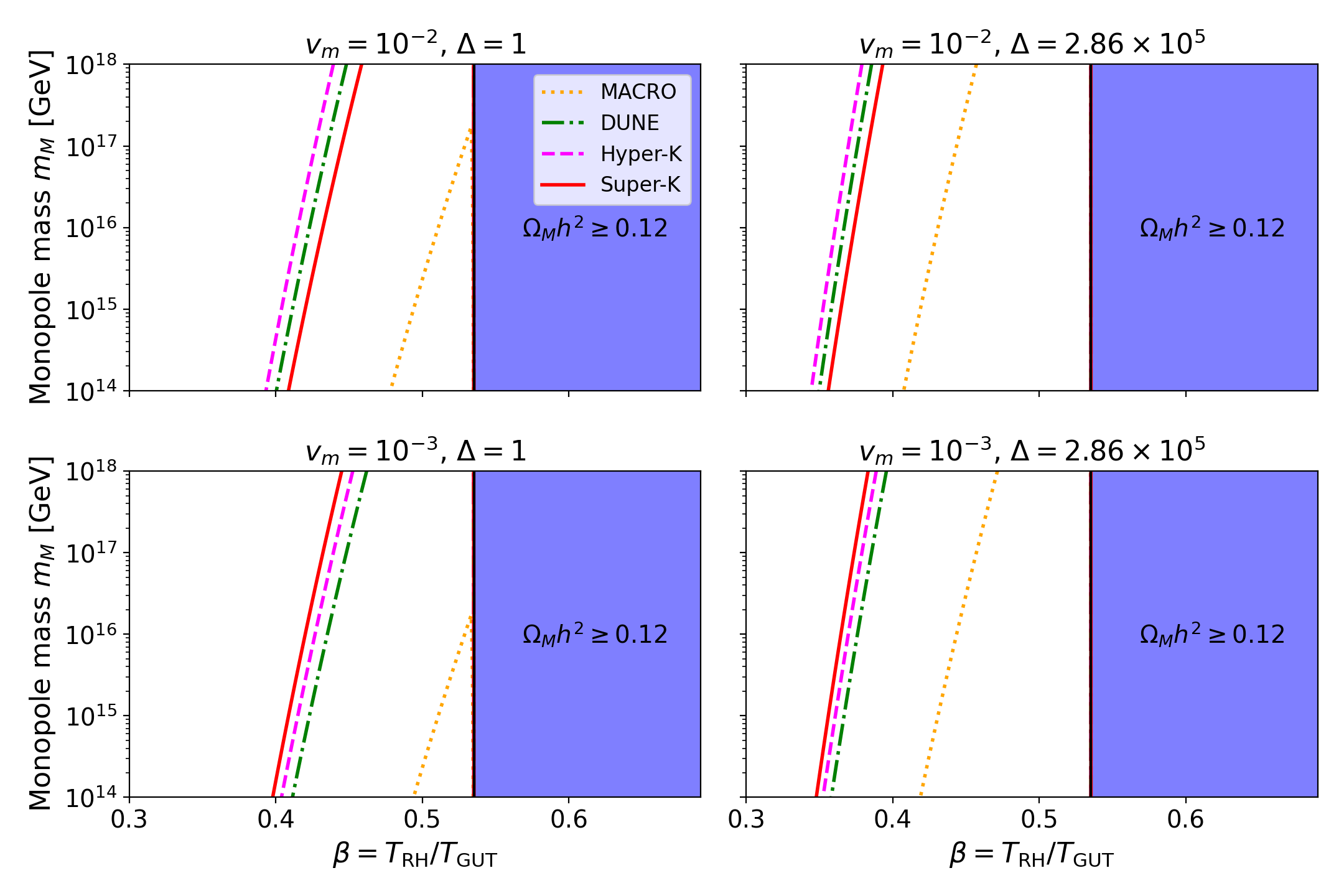}
    \caption{Experimental and DM bounds on monopole abundance as a function of the monopole mass $m_M$ and the normalized reheating temperature $\beta$, for various values of the monopole velocity $v$ and overdensity $\Delta$.}
    \label{fig:beta-obs-exp}
\end{figure}

\begin{figure}
    \centering
    \includegraphics[width=\linewidth]{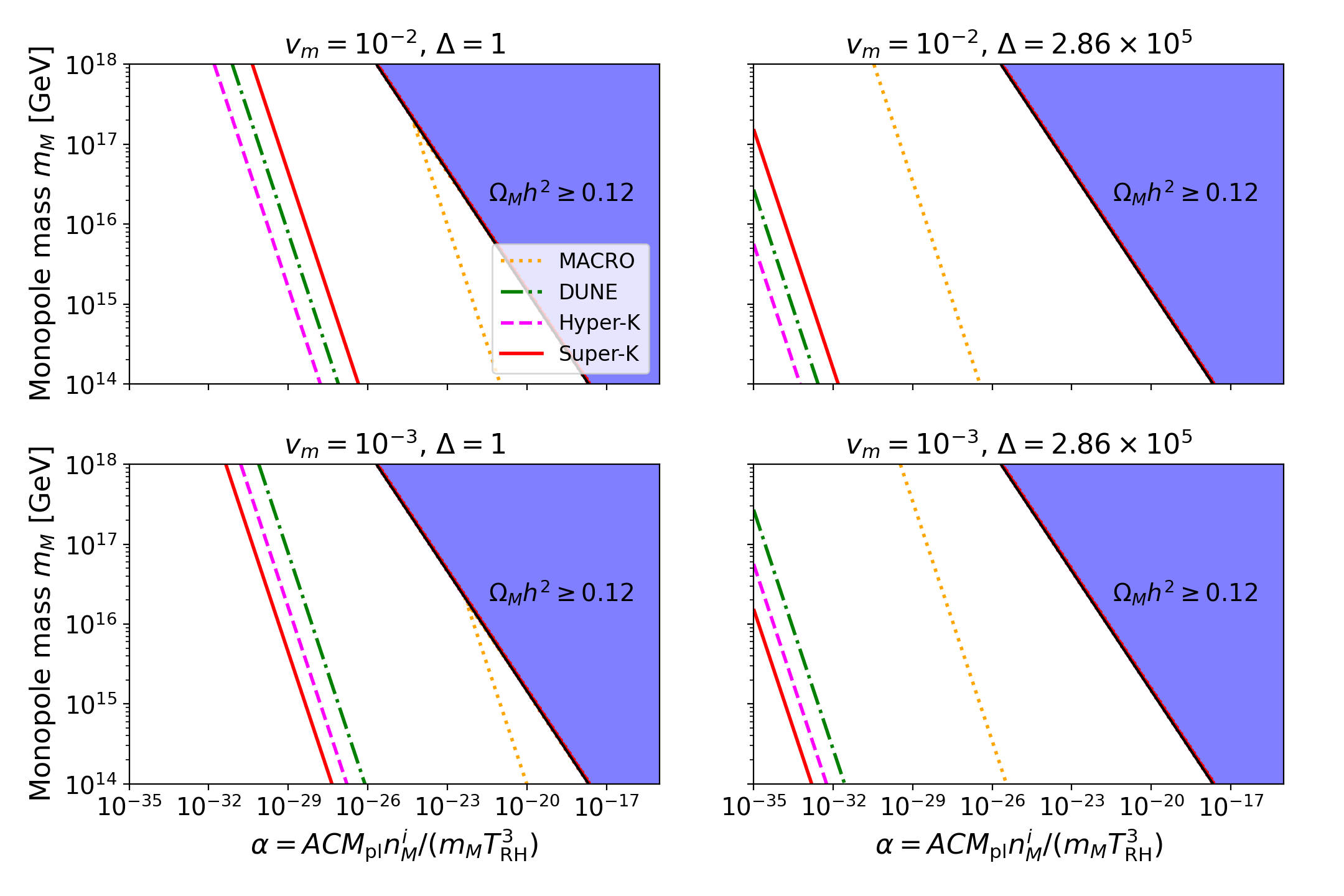}
    \caption{Experimental and DM bounds on monopole abundance as a function of the monopole mass $m_M$ and the normalized initial monopole density $\alpha$, for various values of the monopole velocity $v$ and overdensity $\Delta$.}
    \label{fig:alpha-obs-exp}
\end{figure}

In this section, we summarize the monopole density bounds given by experimental detection: both direct searches for monopoles that have already taken place (MACRO, Super-Kamiokande) and neutrino detection experiments to be performed in the future (DUNE, Hyper-Kamiokande).

The MACRO experiment \cite{MACRO_2002} searched for magnetic monopoles in the velocity range $v_M \in (4\times 10^{-5},1)$. The final results provided the bound $F_M < 1.4\times 10^{-16}\,\text{cm}^{-2}\text{s}^{-1}\text{sr}^{-1}$ on monopole flux. This is $\sim 7\times$ more stringent than the Milky Way Parker bound, but less stringent than the Andromeda Parker bound. Additionally, the Super-Kamiokande collaboration carried out a detailed analysis of proton decay catalysis (Callan-Rubov process) via interaction with GUT monopoles captured by the Sun and reported a bound $F_M \lesssim 6.3\times 10^{-24} (v_M/10^{-3})^2\,\text{cm}^{-2}\text{s}^{-1}\text{sr}^{-1}$, which is substantially more stringent \cite{ueno2012search}. This bound assumes that the cross section of the monopole-catalyzed proton decay in the relativistic limit is $\sigma_0=1\, \rm mb$.

Future experiments will improve bounds on the proton lifetime, and thus, may modify bounds on the flux of magnetic monopoles. This process was described in detail by Candela et al. \cite{Candela_2025}, who predicted the DUNE and Hyper-Kamiokande non-relativistic monopole flux bounds to be $F_M < 1.1\times 10^{-23}\,\text{cm}^{-2}\text{s}^{-1}\text{sr}^{-1}$ and $F_M < 2.3\times 10^{-23}\,\text{cm}^{-2}\text{s}^{-1}\text{sr}^{-1}$ respectively, assuming proton decay is not found.

The translation of these flux bounds into bounds on $\beta$ and $\alpha$ are depicted for a variety of monopole masses, in \Cref{fig:beta-obs-exp} and \Cref{fig:alpha-obs-exp}, respectively. As in the case of the Parker bounds, the most stringent bound occurs when monopoles cluster in a manner similar to DM; in this case, the MACRO and Super-Kamiokande experiment constrains $\beta \lesssim 0.4$ and $\beta \lesssim 0.35$, respectively. Improved proton decay bounds from DUNE and Hyper-Kamiokande are not expected to meaningfully change this result.

\section{Discussion}\label{sec:discussion}

\subsection{Alternative solutions to thermal monopole overclosure}\label{sec:solutions}

Our investigation naturally raises questions about other ways in which the problem of monopole overclosure may be resolved. We observe that since $\Omega_M \sim m_MY_\infty$, thermal monopoles of intermediate mass (in contrast to superheavy GUT-scale monopoles) may lead to a sufficiently small $\Omega_M$ to evade DM bounds. Thus, intermediate-mass monopoles may lead to a non-problematic monopole abundance. However, at this mass scale, monopoles are accelerated to relativistic speeds via Galactic magnetic fields, leading to a different interpretation of the Parker and direct detection bounds. For a review of contemporary bounds on relativistic monopoles, see \cite{Zhang_2024}.

Thermal monopoles may be avoided altogether if their masses are higher than the fiducial value expected of the 't Hooft-Polyakov solution. This is especially possible in the case of high-rank non-abelian subgroups \cite{weinberg1980fundamental}, or if the Higgs self-coupling is extremely large \cite{kirkman1981asymptotic}.

Furthermore, the solution may lie in further symmetry-breaking mechanisms: higher-dimensional topological defects such as cosmic strings and domain walls may facilitate efficient annihilation \cite{Dvali_1998, dunsky2022guts}. A specific example of a topological solution to the monopole problem is the Langacker-Pi mechanism \cite{Langacker_Pi_180}, in which monopoles are connected to each other by flux tubes and self-annihilate. It would be interesting to further study the interaction of strings and domain walls with thermal monopole production, and to find a window of parameters in which interesting stringy dynamics can take place while overclosure is not a problem. In this scenario, gravitational waves produced by the acceleration of monopoles and/or a cosmic string network may be accessible in future experiments \cite{dunsky2022guts}.

Finally, primordial black holes may capture monopoles, as proposed by Stojkovic and Freese \cite{Stojkovic_2005}. This could provide a mechanism to facilitate the annihilation of monopoles which would otherwise be problematic. This solution may also leave a GW signature \cite{yuan2019probing}.

\subsection{Applications to cosmology}
Our analysis has direct implications for cosmology. Specifically, our calculations demand that care must be taken when constructing models which involve a reheating temperature near the GUT scale. In these cases, thermal fluctuations may yield an unacceptably high abundance of monopoles based on observational and experimental constraints. One of the solutions of \Cref{sec:solutions} (or a similar solution) must then be adopted.

Our results also have direct application to monopoles which arise in a hidden sector with gauge group $G_{\rm hid}$ independent of the Standard Model. There has been recent interest in these so-called \textit{dark monopoles} as a DM candidate \cite{khoze2014dark, brummer2026no, brummer2026price}. In a similar vein to \cite{liveoak2026dark}, our model is readily applied to constrain hidden sectors which are reheated near their spontaneous symmetry breaking scale, since particles charged under $G_{\rm hid}$ may thermally produce hidden sector monopoles via the same inverse annihilation process as in the case of GUT monopoles. If the hidden sector contains light charged particles, the bounds derived in \Cref{sec:anal} apply, constraining the maximum temperature of the sector.

\section{Conclusion}\label{sec:conclusion}
In this paper, we studied the thermal production of GUT-scale monopoles in high-temperature reheating scenarios. We found that the present-day abundance of monopoles strongly depends on the ratio $\beta = T_\text{RH}/T_\text{GUT}$. When $\beta \sim O(0.1)$, monopole-antimonopole pairs may be produced thermally from the bath of relativistic charged particles. Our study improved on earlier treatments of this mechanism by retaining the full relativistic form of the equilibrium monopole abundance, incorporating additional BSM degrees of freedom into the monopole annihilation rate, and comparing with modern observational and experimental efforts.

We expressed our results as a bound on the fractional monopole energy density $\Omega_M$. Analytically, when monopoles are produced via reverse annihilation in a diffusive capture process, we obtained $\beta \lesssim 0.55$ for the requirement that $\Omega_M$ does not exceed the energy density of dark matter $\Omega_{\text{DM}}h^2 \sim 0.12$. This bound is independent of the monopole mass, and depends only weakly on the spectrum of charged particles. We reproduced the result numerically, taking into account both diffusive capture and the weaker bremsstrahlung capture, and obtain the consistent bound $\beta \lesssim 0.55$.

Then, we discussed how our bounds on $\beta$ may be further interpreted in the context of astrophysical and experimental bounds on the flux of monopoles. Specifically, we analyzed our reheating bound in the context of Parker bounds, as well as experimental results from MACRO, Super-Kamiokande and the future experiments DUNE and Hyper-Kamiokande. The maximum reheating temperature generally depends on the typical velocity and clustering dynamics of the monopoles, but $\beta \lesssim 0.35$ generally yields an acceptably low abundance.

Several extensions of this analysis would be worthwhile. Intermediate-mass monopoles, which may be accelerated to relativistic speeds by Galactic magnetic fields, have substantially different observational and experimental bounds. In addition, future work should study symmetry-breaking patterns that introduce strings or other topological defects that enhance monopole annihilation. Primordial black holes may also provide an interesting solution to the monopole problem. A more complete treatment of these effects may give insight into whether otherwise excluded reheating temperatures may remain phenomenologically viable.

\begin{acknowledgments}
DL is supported by the National Science Foundation Graduate Research Fellowship Program. JW acknowledges support from the Leinweber Foundation.
\end{acknowledgments}

\bibliography{apssamp}

@article{preskill1979cosmological,
  title={Cosmological production of superheavy magnetic monopoles},
  author={Preskill, John P},
  journal={Physical Review Letters},
  volume={43},
  number={19},
  pages={1365},
  year={1979},
  publisher={APS}
}

@article{ellis1991probing,
  title={Probing the desert using gauge coupling unification},
  author={Ellis, John and Kelley, S and Nanopoulos, Dimitri V},
  journal={Physics Letters B},
  volume={260},
  number={1-2},
  pages={131--137},
  year={1991},
  publisher={Elsevier}
}

@article{pati1974lepton,
  title={Lepton number as the fourth ``color"},
  author={Pati, Jogesh C and Salam, Abdus},
  journal={Physical Review D},
  volume={10},
  number={1},
  pages={275},
  year={1974},
  publisher={APS}
}

@article{georgi1974unity,
  title={Unity of all elementary-particle forces},
  author={Georgi, Howard and Glashow, Sheldon L},
  journal={Physical Review Letters},
  volume={32},
  number={8},
  pages={438},
  year={1974},
  publisher={APS}
}

@article{kolb1996grand,
  title={Grand-unified-theory baryogenesis after preheating},
  author={Kolb, Edward W and Linde, Andrei and Riotto, Antonio},
  journal={Physical Review Letters},
  volume={77},
  number={21},
  pages={4290},
  year={1996},
  publisher={APS}
}

@article{hooper2021gut,
  title={GUT baryogenesis with primordial black holes},
  author={Hooper, Dan and Krnjaic, Gordan},
  journal={Physical Review D},
  volume={103},
  number={4},
  pages={043504},
  year={2021},
  publisher={APS}
}

@incollection{gell2010complex,
  title={Complex spinors and unified theories},
  author={Gell-Mann, Murray and Ramond, Pierre and Slansky, Richard},
  booktitle={Murray Gell-Mann: Selected Papers},
  pages={266--272},
  year={2010},
  publisher={World Scientific}
}

@article{gross1985heterotic,
  title={Heterotic string},
  author={Gross, David J and Harvey, Jeffrey A and Martinec, Emil and Rohm, Ryan},
  journal={Physical Review Letters},
  volume={54},
  number={6},
  pages={502},
  year={1985},
  publisher={APS}
}

@article{yoshimura1978unified,
  title={Unified gauge theories and the baryon number of the universe},
  author={Yoshimura, Motohiko},
  journal={Physical Review Letters},
  volume={41},
  number={5},
  pages={281},
  year={1978},
  publisher={APS}
}

@article{minkowski1977mu,
  title={$\mu$→ e$\gamma$ at a rate of one out of 109 muon decays?},
  author={Minkowski, Peter},
  journal={Physics Letters B},
  volume={67},
  number={4},
  pages={421--428},
  year={1977},
  publisher={Elsevier}
}

@article{guth1981inflationary,
  title={Inflationary universe: A possible solution to the horizon and flatness problems},
  author={Guth, Alan H},
  journal={Physical Review D},
  volume={23},
  number={2},
  pages={347},
  year={1981},
  publisher={APS}
}

@article{turner1982thermal,
  title={Thermal production of superheavy magnetic monopoles in the early universe},
  author={Turner, Michael S},
  journal={Physics Letters B},
  volume={115},
  number={2},
  pages={95--98},
  year={1982},
  publisher={Elsevier}
}

@article{t1974magnetic,
  title={Magnetic monopoles in unified theories},
  author={t Hooft, Gerardus},
  journal={Nucl. Phys. B},
  volume={79},
  number={CERN-TH-1876},
  pages={276--284},
  year={1974},
  publisher={CM-P00060463}
}

@article{polyakov1974particle,
  title={Particle spectrum in quantum field theory},
  author={Polyakov, Alexander M},
  journal={JETP lett},
  volume={20},
  number={194-195},
  pages={300},
  year={1974},
  publisher={World Scientific}
}

@article{zurek1985cosmological,
  title={Cosmological experiments in superfluid helium?},
  author={Zurek, Wojciech H},
  journal={Nature},
  volume={317},
  number={6037},
  pages={505--508},
  year={1985},
  publisher={Nature Publishing Group UK London}
}

@article{kibble1976topology,
  title={Topology of cosmic domains and strings},
  author={Kibble, Thomas WB},
  journal={Journal of Physics A: Mathematical and General},
  volume={9},
  number={8},
  pages={1387--1398},
  year={1976}
}

@article{takenaka2020search,
  title={{Search for proton decay via p→ e+ $\pi$ 0 and p→ $\mu$+ $\pi$ 0 with an enlarged fiducial volume in Super-Kamiokande I-IV}},
  author={Takenaka, A and Abe, K and Bronner, C and Hayato, Y and Ikeda, M and Imaizumi, S and Ito, H and Kameda, J and Kataoka, Y and Kato, Y and others},
  journal={Physical Review D},
  volume={102},
  number={11},
  pages={112011},
  year={2020},
  publisher={APS}
}

@article{OHLSSON2023116268,
title = {Proton decay},
journal = {Nuclear Physics B},
volume = {993},
pages = {116268},
year = {2023},
issn = {0550-3213},
doi = {https://doi.org/10.1016/j.nuclphysb.2023.116268},
url = {https://www.sciencedirect.com/science/article/pii/S0550321323001979},
author = {Tommy Ohlsson},
}

@article{dimopoulos1981softly,
  title={{Softly broken supersymmetry and $SU (5)$}},
  author={Dimopoulos, Savas and Georgi, Howard},
  journal={Nuclear Physics B},
  volume={193},
  number={1},
  pages={150--162},
  year={1981},
  publisher={Elsevier}
}

@article{PhysRevD.32.2348,
  title = {Nucleon decay in supergravity unified theories},
  author = {Nath, Pran and Chamseddine, A. H. and Arnowitt, R.},
  journal = {Phys. Rev. D},
  volume = {32},
  issue = {9},
  pages = {2348--2358},
  numpages = {0},
  year = {1985},
  month = {Nov},
  publisher = {American Physical Society},
  doi = {10.1103/PhysRevD.32.2348},
  url = {https://link.aps.org/doi/10.1103/PhysRevD.32.2348}
}

@article{BABU1998337,
title = {Suggested new modes in supersymmetric proton decay},
journal = {Physics Letters B},
volume = {423},
number = {3},
pages = {337-347},
year = {1998},
issn = {0370-2693},
doi = {https://doi.org/10.1016/S0370-2693(98)00108-7},
url = {https://www.sciencedirect.com/science/article/pii/S0370269398001087},
author = {K.S. Babu and Jogesh C. Pati and Frank Wilczek},
}

@article{SAKAI1982533,
title = {Proton decay in a class of supersymmetric grand unified models},
journal = {Nuclear Physics B},
volume = {197},
number = {3},
pages = {533-542},
year = {1982},
issn = {0550-3213},
doi = {https://doi.org/10.1016/0550-3213(82)90457-6},
url = {https://www.sciencedirect.com/science/article/pii/0550321382904576},
author = {N. Sakai and Tsutomu Yanagida}
}

@article{Parker1970,
  author    = {E. N. Parker},
  title     = {The Origin of Magnetic Fields},
  journal   = {Astrophysical Journal},
  volume    = {160},
  pages     = {383},
  year      = {1970},
  doi       = {10.1086/150442}
}

@article{TurnerParkerBogdan1982,
  author    = {M. S. Turner and E. N. Parker and T. Bogdan},
  title     = {Magnetic monopoles and the survival of galactic magnetic fields},
  journal   = {Physical Review D},
  volume    = {26},
  pages     = {1296},
  year      = {1982}
}

@article{Elyutin:1978recombination,
  author  = {Elyutin, P. V.},
  title   = {Classical Recombination Cross Section},
  journal = {Theoretical and Mathematical Physics},
  volume  = {34},
  number  = {2},
  pages   = {112--115},
  year    = {1978}
}

@article{Zhang_2024,
   title={On the cosmological abundance of magnetic monopoles},
   volume={2024},
   ISSN={1029-8479},
   url={http://dx.doi.org/10.1007/JHEP08(2024)220},
   DOI={10.1007/jhep08(2024)220},
   number={8},
   journal={Journal of High Energy Physics},
   publisher={Springer Science and Business Media LLC},
   author={Zhang, Chen and Zhang, Shi-Hao and Fu, Bowen and Zhang, Jing-Fei and Zhang, Xin},
   year={2024},
   month=Aug}

@article{MACRO_2002,
   title={Final results of magnetic monopole searches with the MACRO experiment},
   volume={25},
   ISSN={1434-6052},
   url={http://dx.doi.org/10.1140/epjc/s2002-01046-9},
   DOI={10.1140/epjc/s2002-01046-9},
   number={4},
   journal={The European Physical Journal C},
   publisher={Springer Science and Business Media LLC},
   author={ and Ambrosio et al., M.},
   year={2002},
   month=Nov, pages={511–522} }

@article{Particle_Data_Review_2022,
    author = {{Particle Data Group} and Workman, R L et. al.},
    title = {Review of Particle Physics},
    journal = {Progress of Theoretical and Experimental Physics},
    volume = {2022},
    number = {8},
    pages = {083C01},
    year = {2022},
    month = {08},
    issn = {2050-3911},
    doi = {10.1093/ptep/ptac097},
    url = {https://doi.org/10.1093/ptep/ptac097},
    eprint = {https://academic.oup.com/ptep/article-pdf/2022/8/083C01/49175539/ptac097.pdf},
}

@article{Medvedev_2017,
   title={Plasma constraints on the cosmological abundance of magnetic monopoles and the origin of cosmic magnetic fields},
   volume={2017},
   ISSN={1475-7516},
   url={http://dx.doi.org/10.1088/1475-7516/2017/06/058},
   DOI={10.1088/1475-7516/2017/06/058},
   number={06},
   journal={Journal of Cosmology and Astroparticle Physics},
   publisher={IOP Publishing},
   author={Medvedev, Mikhail V. and Loeb, Abraham},
   year={2017},
   month=06, pages={058–058} }

@article{Candela_2025,
   title={Monopoles at future neutrino detectors},
   volume={2025},
   ISSN={1029-8479},
   url={http://dx.doi.org/10.1007/JHEP07(2025)034},
   DOI={10.1007/jhep07(2025)034},
   number={7},
   journal={Journal of High Energy Physics},
   publisher={Springer Science and Business Media LLC},
   author={Candela, Pablo M. and Khoze, Valentin V. and Turner, Jessica},
   year={2025},
   month=07 }

@article{dunsky2022guts,
  title={GUTs, hybrid topological defects, and gravitational waves},
  author={Dunsky, David I and Ghoshal, Anish and Murayama, Hitoshi and Sakakihara, Yuki and White, Graham},
  journal={Physical Review D},
  volume={106},
  number={7},
  pages={075030},
  year={2022},
  publisher={APS}
}

@article{Dvali_1998,
  title = {Sweeping Away the Monopole Problem},
  author = {Dvali, G. and Liu, Hong and Vachaspati, Tanmay},
  journal = {Phys. Rev. Lett.},
  volume = {80},
  issue = {11},
  pages = {2281--2284},
  numpages = {0},
  year = {1998},
  month = {Mar},
  publisher = {American Physical Society},
  doi = {10.1103/PhysRevLett.80.2281},
  url = {https://link.aps.org/doi/10.1103/PhysRevLett.80.2281}
}

@article{Langacker_Pi_180,
  title = {Magnetic Monopoles in Grand Unified Theories},
  author = {Langacker, Paul and Pi, So-Young},
  journal = {Phys. Rev. Lett.},
  volume = {45},
  issue = {1},
  pages = {1--4},
  numpages = {0},
  year = {1980},
  month = {Jul},
  publisher = {American Physical Society},
  doi = {10.1103/PhysRevLett.45.1},
  url = {https://link.aps.org/doi/10.1103/PhysRevLett.45.1}
}

@article{Stojkovic_2005,
   title={A black hole solution to the cosmological monopole problem},
   volume={606},
   ISSN={0370-2693},
   url={http://dx.doi.org/10.1016/j.physletb.2004.12.019},
   DOI={10.1016/j.physletb.2004.12.019},
   number={3-4},
   journal={Physics Letters B},
   publisher={Elsevier BV},
   author={Stojkovic, Dejan and Freese, Katherine},
   year={2005},
   month=Jan, pages={251–257} }

@book{vilenkin1994cosmic,
  title={Cosmic strings and other topological defects},
  author={Vilenkin, Alexander and Shellard, E Paul S},
  year={1994},
  publisher={Cambridge university press}
}

@article{brummer2026no,
  title={No room for minimal monopole dark matter},
  author={Br{\"u}mmer, Felix and Ferrante, Giacomo and Fischer, Th{\'e}odore and Frigerio, Michele},
  journal={Physical Review D},
  volume={113},
  number={9},
  pages={L091701},
  year={2026},
  publisher={APS}
}

@article{prasad1975exact,
  title={{Exact classical solution for the't Hooft monopole and the Julia-Zee dyon}},
  author={Prasad, Manoj K and Sommerfield, Charles M},
  journal={Physical Review Letters},
  volume={35},
  number={12},
  pages={760},
  year={1975},
  publisher={APS}
}

@article{planck2020planck,
  title={{Planck 2018 results: VI. Cosmological parameters}},
  journal={Astronomy \& Astrophysics},
  volume={641},
  pages={A6},
  year={2020},
  publisher={EDP Sciences}
}

@article{sofue2020rotation,
  title={{Rotation curve of the Milky Way and the dark matter density}},
  author={Sofue, Yoshiaki},
  journal={Galaxies},
  volume={8},
  number={2},
  pages={37},
  year={2020},
  publisher={MDPI}
}

@article{turner1982magnetic,
  title={Magnetic monopoles and the survival of galactic magnetic fields},
  author={Turner, Michael S and Parker, Eugene N and Bogdan, TJ},
  journal={Physical Review D},
  volume={26},
  number={6},
  pages={1296},
  year={1982},
  publisher={APS}
}

@article{weinberg1980fundamental,
  title={Fundamental monopoles and multimonopole solutions for arbitrary simple gauge groups},
  author={Weinberg, Erick J},
  journal={Nuclear Physics B},
  volume={167},
  number={3},
  pages={500--524},
  year={1980},
  publisher={Elsevier}
}

@article{khoze2014dark,
  title={Dark matter monopoles, vectors and photons},
  author={Khoze, Valentin V and Ro, Gunnar},
  journal={Journal of High Energy Physics},
  volume={2014},
  number={10},
  pages={1--27},
  year={2014},
  publisher={Springer}
}

@article{brummer2026price,
  title={The price for monopole dark matter},
  author={Br{\"u}mmer, Felix and Ferrante, Giacomo and Fischer, Th{\'e}odore and Frigerio, Michele},
  journal={arXiv preprint arXiv:2607.29492},
  year={2026}
}

@article{liveoak2026dark,
  title={{Dark Monopoles, Bounds on Hidden Sectors, and Cosmological Implications}},
  author={Liveoak, Donald and Maharana, Anshuman and Wells, James D},
  journal={arXiv preprint arXiv:2607.20843},
  year={2026}
}

@article{yuan2019probing,
  title={Probing primordial--black-hole dark matter with scalar induced gravitational waves},
  author={Yuan, Chen and Chen, Zu-Cheng and Huang, Qing-Guo},
  journal={Physical Review D},
  volume={100},
  number={8},
  pages={081301},
  year={2019},
  publisher={APS}
}

@article{kirkman1981asymptotic,
  title={Asymptotic analysis of the monopole structure},
  author={Kirkman, Thomas W and Zachos, Cosmas K},
  journal={Physical Review D},
  volume={24},
  number={4},
  pages={999},
  year={1981},
  publisher={APS}
}

@article{zeldovich1978concentration,
  title={On the concentration of relic magnetic monopoles in the universe},
  author={Zeldovich, Ya B and Khlopov, M Yu},
  journal={Physics Letters B},
  volume={79},
  number={3},
  pages={239--241},
  year={1978},
  publisher={Elsevier}
}

@article{kolb1981early,
  title={The early universe},
  author={Kolb, Edward W and Turner, Michael S},
  journal={Nature},
  volume={294},
  number={5841},
  pages={521--526},
  year={1981},
  publisher={Nature Publishing Group UK London}
}

@article{ueno2012search,
  title={{Search for GUT monopoles at Super--Kamiokande}},
  author={Ueno, Koh and Abe, Kou and Hayato, Yoshinari and Iida, Takashi and Iyogi, Kazuki and Kameda, Jun and Koshio, Yusuke and Kozuma, Y and Miura, Makoto and Moriyama, Shigetaka and others},
  journal={Astroparticle Physics},
  volume={36},
  number={1},
  pages={131--136},
  year={2012},
  publisher={Elsevier}
}

@article{collins1984thermal,
  title={Thermal production of superheavy magnetic monopoles in the new inflationary-Universe scenario},
  author={Collins, William and Turner, Michael S},
  journal={Physical Review D},
  volume={29},
  number={10},
  pages={2158},
  year={1984},
  publisher={APS}
}

\end{document}